\documentclass[twocolumn]{aastex631}

\PassOptionsToPackage{table}{xcolor}
\usepackage[table]{xcolor}
\usepackage{tabularx,colortbl}
\usepackage{graphicx}
\usepackage{color,colortbl}
\usepackage{amsmath}
\usepackage{enumitem}
\usepackage{multirow}

\def\apjl{ApJL}               % Astrophysical Journal, Letters
\def\apjs{ApJS}               % Astrophysical Journal, Supplement
 
\def\aap{A\&A}                % Astronomy and Astrophysics
\definecolor{fuchsia}{rgb}{1.0, 0.0, 1.0}

\definecolor{byzantine}{rgb}{0.74, 0.2, 0.64}

\definecolor{pink}{RGB}{255, 20, 147}

\definecolor{amber(sae/ece)}{rgb}{1.0, 0.49, 0.0}

\definecolor{carminered}{rgb}{1.0, 0.0, 0.22}

\definecolor{bittersweet}{rgb}{1.0, 0.44, 0.37}

\definecolor{ao(english)}{rgb}{0.0, 0.5, 0.0}

\newcommand{\add}[1]{{\color{black} #1}}
\newcommand{\addd}[1]{{\color{black} #1}}

\begin{document}

\title{\bf The IceCube Pie Chart: Relative Source Contributions to the Cosmic Neutrino Flux}
\author{I. Bartos}
\thanks{imrebartos@ufl.edu}
\affiliation{Department of Physics, University of Florida, PO Box 118440, Gainesville, FL 32611-8440, USA}
\author{D. Veske}
\affiliation{Department of Physics, Columbia University, New York, NY 10027, USA}
\author{M. Kowalski}
\affiliation{Deutsches Elektronen Synchrotron DESY, Platanenallee 6, 15738 Zeuthen, Germany}
\affiliation{Institut fur Physik, Humboldt-Universit\:at zu Berlin, D-12489 Berlin, Germany}
\affiliation{Columbia Astrophysics Laboratory, Columbia University, New York, NY 10027, USA}
\author{Z. M\'arka}
\affiliation{Columbia Astrophysics Laboratory, Columbia University, New York, NY 10027, USA}
\author{S. M\'arka}
\affiliation{Department of Physics, Columbia University, New York, NY 10027, USA}

\begin{abstract}
Neutrino events from IceCube have recently been associated with multiple astrophysical sources. Interestingly, these likely detections represent three distinct astrophysical source types: active galactic nuclei (AGN), blazars, and tidal disruption events (TDE). Here we compute the expected contributions of AGNs, blazars and TDEs to the overall cosmic neutrino flux detected by IceCube based on the associated events, IceCube's sensitivity, and the source types' astrophysical properties. We find that, despite being the most commonly identified sources, blazars cannot contribute more than 11\% of the total flux (90\% credible level), consistent with existing limits from stacked searches. On the other hand, we find that either AGNs or TDEs could contribute more than 50\% of the total flux (90\% credible level), \add{although stacked searches further limit the TDE contribution to $\lesssim 30\%$.} We also find that so-far unknown source types contribute at least 10\% of the total cosmic flux with a probability of 80\%. We assemble a pie chart that shows the most likely fractional contribution of each source type to IceCube's total neutrino flux. 
\vspace{1cm}
\end{abstract}

%%%%%%%%%%%%%%%%%%%%%%%%%%%%%%%%%%%%%%%%%%%%%%%%%%%%
\section{Introduction}
%%%%%%%%%%%%%%%%%%%%%%%%%%%%%%%%%%%%%%%%%%%%%%%%%%%%

The Universe produces a quasi-diffuse flux of high-energy ($>\,$TeV) neutrinos (hereafter cosmic neutrino flux; \citealt{2013Sci...342E...1I,2014PhRvL.113j1101A}), whose properties are now well characterized \citep{2016ApJ...833....3A}. The origin of this cosmic flux is, nevertheless, not yet understood. 

High-energy neutrinos have recently been identified from several distinct astrophysical sites. These sources include the TXS\,0506+056 blazar \citep{2018Sci...361.1378I,2018Sci...361..147I}, a nearby Seyfert galaxy (NGC\,1068; \citealt{2020PhRvL.124e1103A}), and a tidal disruption event (TDE; AT2019dsg; \citealt{2021NatAs...5..510S}). Several other blazars have also been identified with probable high-energy neutrino associations \citep{2016NatPh..12..807K,2020arXiv200909792K}. At the same time, \add{the contribution of several source types to the overall flux have been constrained, including  blazars \citep{2020ApJ...890...25Y}, and TDEs \citep{2017ApJ...838....3S,2020ApJ...902..108M,2016PhRvD..94j3006M,2018A&A...616A.179G, 2019ICRC...36.1016S} and gamma-ray bursts (GRBs; \citealt{2012Natur.484..351I}) }.

In this paper we evaluate the expected overall contribution of different source types to the cosmic neutrino flux based on the \add{associated {\it individual} neutrino sources, taking into account the uncertainty in source associations.} We show that the widely different properties of different source types mean that the same number of detections translates to different expected contributions to the overall flux. In addition, computing the expected contribution of each source type to the overall cosmic neutrino flux allows us to estimate the fraction of the overall flux that arrives from so far unidentified source types, i.e sources that are not AGNs, blazars or TDEs.

An additional goal of the present description is to demonstrate how different source features, such as their number density, number of detections or their cosmic rate evolution, contributes to their estimated contribution to the overall neutrino flux. Therefore, before discussing our Bayesian estimate that includes most accessible details about the source populations, we first compute fractional contributions through simpler estimates in which the role of different source properties is more accessible. 

The paper is organized as follows.  We first carry out our simplest "warm-up" calculation of fractional contributions in Section \ref{sec:warmup} using detections that followed high-energy neutrino alerts publicly released by IceCube. Next, we introduce a simplified model in Section \ref{sec:simplemodel} that highlights the relative importance of population properties. We then introduce our most detailed and realistic method in Section \ref{sec:fullmodel} and its implementation in \ref{sec:implementation} to obtain the estimated relative contributions for both detected and unknown source types. Results for this most detailed model are presented in Section \ref{sec:results}.  We summarize our conclusions in Section \ref{sec:conclusion}.

%%%%%%%%%%%%%%%%%%%%%%%%%%%%%%%%%%%%%%%%%%%%%%%%%%%%
\section{Warming up: the fraction of discoveries} \label{sec:warmup}
%%%%%%%%%%%%%%%%%%%%%%%%%%%%%%%%%%%%%%%%%%%%%%%%%%%%

It is instructive to evaluate the fraction of high-energy neutrino alerts \citep{2017APh....92...30A,2019ICRC...36.1021B} that lead to likely associations with counterparts. If one further factors in the completness of the catalog of potential counterparts, one can translate this fraction into a constraint on the contribution of the specific source class to the total neutrino flux.  Below we examine blazars and TDEs which were identified so far in association with IceCube's neutrino alerts.

We first consider blazars that were found in association with the $N_{\rm alert} \sim60$ alerts\footnote{\url{https://gcn.gsfc.nasa.gov/amon_hese_events.html}.}\footnote{\url{https://gcn.gsfc.nasa.gov/amon_ehe_events.html}.} sent to the community by IceCube so far: besides the well known TXS0506+056 \citep{2018Sci...361..147I}, more recently PKS 1502+106---another exceptionally bright blazar---was identified in coincident with a well localized neutrino track (\cite{2019ATel12967....1T}). Other claimed associations are less significant (e.g.\ GB6 J1040+0617, \cite{Aartsen:2019gxs}) and/or have not been found in associations with alerts (PKS\,1424-41; \citealt{2016NatPh..12..807K}), and since we do not consider backgrounds here, they are not counted in this simple exercise. We hence count two detections, $N_{\rm det}=2$. We take into account that roughly half of IceCube's alerts are estimated to be of astrophysical origin (signalness\footnote{IceCube neutrino alerts have an assigned quantity called signalness, defined as the probability that the neutrino signal came from an astrophysical, as opposed to an atmospheric source, based on the neutrino energy, sky location and time of arrival, but without accounting for information from electromagnetic observations \cite{2019ICRC...36.1021B}.} $s\approx 0.5$). In addition, based on their gamma-ray flux, we assume that a fraction $f\sim0.7$ of the high-energy neutrino flux from this source type are from electromagnetically resolved blazars \citep{2016PhRvL.116o1105A}. With these quantities we can estimate the fraction of cosmic, high-energy neutrinos due to a blazar population: $N_{\rm det}/N_{\rm alert}/f/s = 0.10$.

Likewise for TDEs, \cite{2021NatAs...5..510S} associated one TDE ($N_{\rm det}=1$) from a search of alerts ($N_{\rm alerts}=9$). \add{We estimate TDEs' completeness factor to be $f=0.5$ that assumes that TDEs can be detected electromagnetically out to $\sim1\,$Gpc}. These numbers translate to a fraction of the diffuse neutrino flux due to a population of TDEs of $N_{\rm det}/N_{\rm alert}/f/s = 0.45$. 

These initial estimates come with significant caveats, e.g. they do not take into account the significance of the observations, and moreover are not generalizable to observations that are not based on single-neutrino coincidences. Nevertheless, they are illustrative of a key point, namely the importance of the completeness of the catalog of counterparts in extrapolating the flux to a full population of sources. 

%%%%%%%%%%%%%%%%%%%%%%%%%%%%%%%%%%%%%%%%%%%%%%%%%%%%
\section{Simple model} \label{sec:simplemodel}
%%%%%%%%%%%%%%%%%%%%%%%%%%%%%%%%%%%%%%%%%%%%%%%%%%%%

In order to characterize the role of different population properties in connecting detections with the estimated contribution to the overall cosmic neutrino flux, we consider the following simple model. We separately consider continuous neutrino sources that are detected through time-integrated searches, and searches that identify sources via the detection of a single high-energy neutrino. \add{Similar computations of converting resolved sources to overall cosmic flux can be found in the literature, see e.g. \cite{2008PhRvD..78h3011L,2010PhRvD..81b3001S,2014PhRvD..90d3005A,2016PhRvD..94j3006M}.}

%%%%%%%%%%%%%%%%%%%%%%%%%%%%%%%%%%%%%%%%%%%%%%%%%%%%%%%%%
\subsection{Time-integrated detection}
%%%%%%%%%%%%%%%%%%%%%%%%%%%%%%%%%%%%%%%%%%%%%%%%%%%%%%%%%

Time-integrated searches are particularly relevant for continuous sources with high number densities where the detection of multiple neutrinos is necessary to claim detection. Let $F_{\rm \nu,0}$ be the neutrino threshold above which such a time-integrated search is able to identify a neutrino flux. For simplicity we neglect any dependence on the sources' neutrino spectrum and sky location.

Let sources within a given source type be uniformly distributed in the local universe with number density $\rho$. Let each source have unknown identical neutrino luminosity $L_\nu$. The maximum distance $r_{\rm max}$ within which these sources can be detected is 
\begin{equation}
r_{\rm max} = \sqrt{\frac{L_\nu}{4\pi F_{\rm \nu,0}}}
\label{eq:rmax}
\end{equation}
Within this distance the expected number of sources is
\begin{equation}
\langle N_{\rm det}\rangle = \frac{4}{3}\pi r_{\rm max}^{3} \rho
\label{eq:Ndet}
\end{equation}
We will consider the number $N_{\rm det}$ of detected sources to be the best estimate of the expected number of detections, i.e. $\langle N_{\rm det}\rangle=N_{\rm det}$. With this, the expected neutrino flux from the considered source type within $r_{\rm max}$ is
\begin{equation}
F_{\rm \nu,r} = \int_{0}^{r_{\rm max}}dr 4\pi r^2 \rho \frac{L_\nu}{4\pi r^2} = \rho L_\nu r_{\rm max} = 3 F_{\rm \nu,0} N_{\rm det},
\end{equation}
where we made use of Eqs. \ref{eq:rmax} and \ref{eq:Ndet} to change variables.

Let $f_{\rm r}(r_{\rm max})$ be the fraction of the total neutrino flux from a given source type that comes from sources within distance $r_{\rm max}$. For not too large distances we approximate \add{the dependence of $f_{\rm r}$ on $r_{\rm max}$ to be linear, i.e. $f_{\rm r}(r_{\rm max})= f_0 r_{\rm max}$, where $f_0$ is a source-type-dependent constant. To compute $f_0$, we need to integrate the neutrino flux from the entire source population accounting for cosmic evolution \citep{2017PhRvD..96b3003B}.}  With this, we can express the overall flux expected from the considered source type as (see also \citealt{2018ApJ...865..124M,2012JCAP...08..030M})
\begin{equation}
F_{\rm \nu,tot}^{\rm (int)} = \frac{F_{\rm \nu,r}(r_{\rm max})}{f_{\rm r}(r_{\rm max})} = \frac{3 F_{\rm \nu,0} N_{\rm det}}{f_0 r_{\rm max}} \approx 5  F_{\rm \nu,0} f_{0}^{-1} N_{\rm det}^{2/3} \rho^{1/3}
\label{eq:ftotint}
\end{equation}
where the superscript "(int)" refers to time-integrated detection. We see that, for the same number of detections, the corresponding flux can be very different due to its dependence on $\rho$, which can vary orders of magnitude between sources.

%%%%%%%%%%%%%%%%%%%%%%%%%%%%%%%%%%%%%%%%%%%%%%%%%%%%%%%%%
\subsection{Single-neutrino detection}
%%%%%%%%%%%%%%%%%%%%%%%%%%%%%%%%%%%%%%%%%%%%%%%%%%%%%%%%%

For rare and/or transient sources, a single high-energy neutrino can be sufficient to claim detection. Let $F_{\rm \nu,1}$ be the source flux for which the {\it expected} number of detected neutrinos is $1$ during the observing period $T_{\rm obs}$. For transient sources with duration $\tau$ the corresponding actual source flux is $T_{\rm obs}/\tau$ times higher.

Let sources within a given source type be uniformly distributed in the local universe with number density $\rho$. Let each source have identical neutrino luminosity $L_\nu$. For transient sources, we will adopt the same notation defining $\rho=\mathcal{R}T_{\rm obs}$ with source rate density $\mathcal{R}$, and $L_\nu=E_{\nu}T_{\rm obs}^{-1}$ with $E_{\nu}$ total energy radiated in high-energy neutrinos.

A single neutrino can only lead to discovery if it is directionally (and temporally for transients) coincident with a source candidate. A single neutrino can be detected with non-vanishing probability even from distant sources. Therefore, single-neutrino searches will not be limited by a distance threshold for detection as in the time-integrated case. Instead, a solution is that the number of source candidates considered in the search is capped in order to limit the search's trial factor.  Source candidates will be selected for a search such that those expected to have the highest neutrino flux at Earth are included. If all sources produce the same neutrino flux then the closest sources are selected. If we limit the source candidates in a search to the number $N_{\rm max}$, then this corresponds to an effective maximum search distance of
\begin{equation}
r_{\rm max}= \left(\frac{3N_{\rm max}}{4\pi \rho}\right)^{1/3}
\label{eq:rmaxsingle}
\end{equation}
The expected number of detected sources within this distance is
\begin{equation}
\langle N_{\rm det}\rangle = \int_{0}^{r_{\rm max}}dr 4\pi r^2 \rho \frac{L_\nu}{4\pi r^2}F_{\rm \nu,1}^{-1} = \rho L_\nu r_{\rm max} F_{\rm \nu,1}^{-1}
\label{eq:Ndetsingle}
\end{equation}
where we assumed that the expected number of detected neutrinos from a single source is $\ll1$. 

We will consider the number $N_{\rm det}$ of detected sources to be the best estimate of the expected number of detections, i.e. $\langle N_{\rm det}\rangle=N_{\rm det}$. With this, the expected neutrino flux from the considered source type within $r_{\rm max}$ is
\begin{equation}
F_{\rm \nu,r}(r_{\rm max}) = \int_{0}^{r_{\rm max}}dr 4\pi r^2 \rho \frac{L_\nu}{4\pi r^2} = \rho L_\nu r_{\rm max} = F_{\rm \nu,1} N_{\rm det},
\end{equation}
where we made use of Eq. \ref{eq:Ndetsingle} to change variables. 

Similarly to the previous section, we introduce the fraction $f_{\rm r}(r_{\rm max})= f_0 r_{\rm max}$ of the total neutrino flux from a given source type that comes from sources within distance $r_{\rm max}$. With this, we can express the overall flux expected from the considered source type as
\begin{equation}
F_{\rm \nu,tot}^{\rm (single)} = \frac{F_{\rm \nu,r}(r_{\rm max})}{f_{\rm r}(r_{\rm max})} = \frac{F_{\rm \nu,1} N_{\rm det}}{f_0 r_{\rm max}} \approx \frac{1.6  F_{\rm \nu,1} N_{\rm det} \rho^{1/3} }{f_{0} N_{\rm max}^{1/3}}
\label{eq:ftotsingle}
\end{equation}
where the superscript "(single)" refers to single-neutrino detection. This formula is very similar to the one we obtained for time integrated searches (see Eq. \ref{eq:ftotint}), and shows that the number density (or rate density) of sources can substantially alter the expected contribution from a source type.

%%%%%%%%%%%%%%%%%%%%%%%%%%%%%%%%%%%%%%%%%%%%%%%%%%%%%%%%%
\subsection{Estimated contribution based on simple model}
%%%%%%%%%%%%%%%%%%%%%%%%%%%%%%%%%%%%%%%%%%%%%%%%%%%%%%%%%

We now estimate the contribution of different detected source types to the cosmic neutrino flux. We additionally consider sources that have so far been undetected to gauge how strong constraint their non-detection represents to their contribution to the overall flux. \add{In this section we made no constraint on the combined contribution of the considered source type, i.e. it is interesting that their total contribution is less then, but comparable to, the total IceCube flux.}

%%%%%%%%%%%%%%%%%%%%%%%%%%%%%%%%%%%%%%%%%%%%%%%%%%%%%%%%%
\subsection{Flux fraction}
%%%%%%%%%%%%%%%%%%%%%%%%%%%%%%%%%%%%%%%%%%%%%%%%%%%%%%%%%

\begin{figure}
\includegraphics[width=0.48\textwidth]{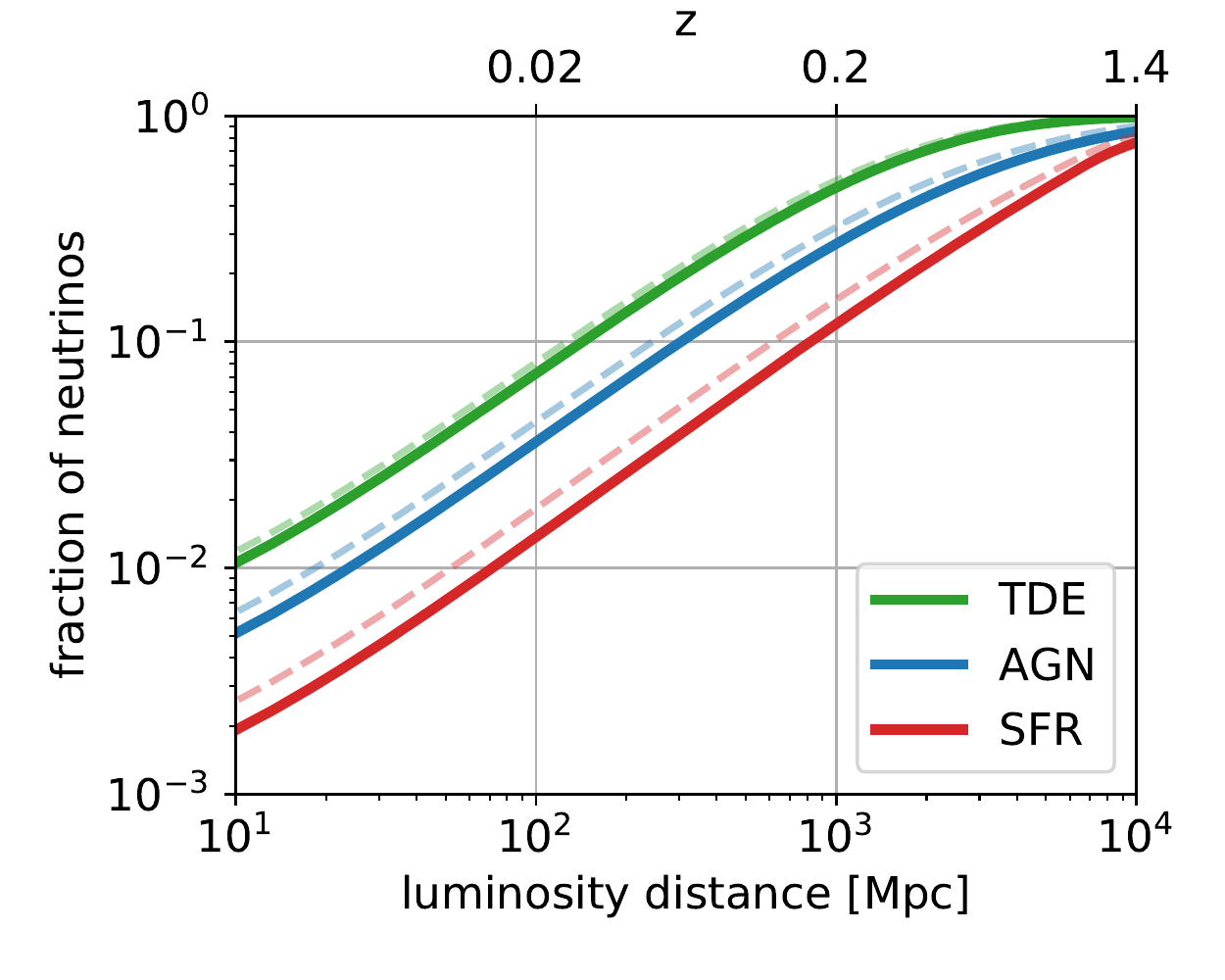}
\centering
\caption{Fraction of detected high-energy neutrinos in IceCube expected from sources closer than luminosity distance $d_{\rm L}$ as a function of $d_{\rm L}$. Shown are results for AGNs/blazars, TDEs and source populations that track the cosmic star formation rate (SFR; see legend). Solid lines show results for assumed neutrino spectral density $dN/dE\propto E^{-2.5}$. For comparison we show for $dN/dE\propto E^{-3}$ (dashed lines).}
\label{fig:neutrinoCDF}
\end{figure}

First, we computed $f_0$ for different source populations, which is needed to estimate their total flux. We considered (non-jetted) AGNs, tidal disruption events (TDEs) and, for comparison, populations that follow the cosmic star-formation rate. Our results are shown in Fig. \ref{fig:neutrinoCDF}. For AGNs we adopted their number density for bolometric luminosity $>10^{43}$\,erg\,s$^{-1}$ from \cite{2015ApJ...802..102L}. For TDEs we adopted their cosmic density evolution from \cite{2015ApJ...812...33S}, while we adopted the cosmic star formation rate from \cite{2008MNRAS.388.1487L}. 

%%%%%%%%%%%%%%%%%%%%%%%%%%%%%%%%%%%%%%%%%%%%%%%%%%%%%%%%%
\subsubsection{Blazars}
%%%%%%%%%%%%%%%%%%%%%%%%%%%%%%%%%%%%%%%%%%%%%%%%%%%%%%%%%

\addd{We adopted the density $\rho_{\rm blazar} = 10^{-9}$\,Mpc$^{-3}$ corresponding to a high-lumminosity class of blazars called flat-spectrum radio quasars (FSRQs; \citealt{2016PhRvD..94j3006M}). With this choice we assume that a subclass of blazars, FSRQs, are mainly responsible for neutrino emission from this class.} We used $N_{\rm det}=3$ and $T_{\rm obs}=5$\,yr. \add{The choice of $T_{\rm obs}$ reflects the duration since IceCube began real-time alerts.} We considered blazars to be detectable with single-neutrino searches. We adopted $\mathcal{F}_{\rm \nu,1}=4\times10^{-10}$\,erg\,cm$^{-2}$\,s$^{-1}(1\mbox{yr}/T_{\rm obs})$, where we assumed a neutrino spectral index of \add{2.5} and neutrino energy range $\varepsilon_\nu\in[100\mbox{GeV},100\mbox{PeV}]$.\footnote{Here we used the flux corresponding to an expected number of 1 detected neutrino in the $\varepsilon_\nu\in[100\mbox{TeV},100\mbox{PeV}]$ from \cite{AMONEHEalert}. We then extrapolated this flux to the energy range $\varepsilon_\nu\in[100\mbox{GeV},100\mbox{PeV}]$ to make it compatible with the flux from the time-integrated search. We adopted a neutrino spectral index of \add{2.5} for the extrapolation.} We further adopted $f_0=5\times10^{-4}$\,Mpc$^{-1}$ (see Fig. \ref{fig:neutrinoCDF}, we considered the same fraction for AGNs and blazars) and $N_{\rm max} = 100$ (see \cite{2020PhRvL.124e1103A}). With these parameters, we obtained \add{$F_{\rm \nu,tot}^{\rm (blazar)} = 3 \times 10^{-9}$\,erg\,cm$^{-2}$\,s$^{-1}$}.

%%%%%%%%%%%%%%%%%%%%%%%%%%%%%%%%%%%%%%%%%%%%%%%%%%%%%%%%%
\subsubsection{AGNs}
%%%%%%%%%%%%%%%%%%%%%%%%%%%%%%%%%%%%%%%%%%%%%%%%%%%%%%%%%

For AGNs we used $N_{\rm det}=1$ and $\rho_{\rm AGN} = 10^{-4}$\,Mpc$^{-3}$. We considered AGNs to be detectable with a time-integrated search. We adopted $F_{\rm \nu,0}=2\times10^{-11}$\,erg\,cm$^{-2}$\,s$^{-1}$ where we assumed a neutrino spectral index of \add{2.5} and energy range $\varepsilon_\nu\in[100\mbox{GeV},100\mbox{PeV}]$. We further adopted $f_0=3\times10^{-4}$\,Mpc$^{-1}$.

With these parameters, we obtained \add{$F_{\rm \nu,tot}^{\rm (AGN)} = 2\times10^{-8}$\,erg\,cm$^{-2}$\,s$^{-1}$.}

%%%%%%%%%%%%%%%%%%%%%%%%%%%%%%%%%%%%%%%%%%%%%%%%%%%%%%%%%
\subsubsection{TDEs}
%%%%%%%%%%%%%%%%%%%%%%%%%%%%%%%%%%%%%%%%%%%%%%%%%%%%%%%%%

For TDEs we used $N_{\rm det}=1$ and $\rho_{\rm TDE} = \mathcal{R}_{\rm TDE}T_{\rm obs} = 10^{-9.5}$\,Mpc$^{-3}$ where we assumed $T_{\rm obs}=5$\,yr. We considered TDEs to be detectable with single-neutrino searches. We adopted $\mathcal{F}_{\rm \nu,1}=4\times10^{-10}$\,erg\,cm$^{-2}$\,s$^{-1}(1\mbox{yr}/T_{\rm obs})$, assuming a neutrino spectral index of 2.5 and neutrino energy range $\varepsilon_\nu\in[100\mbox{GeV},100\mbox{PeV}]$. We adopted $f_0=9\times10^{-4}$\,Mpc$^{-1}$.

In order to determine $N_{\rm max}$, we set the maximum detectable distance of TDEs to be $r_{\rm max,TDE}=1$\,Gpc given the distance range of past identified TDEs \citep{2020arXiv200101409V}. We assumed that electromagnetic follow-up of $\gtrsim100$\,TeV neutrinos could identify every TDE within $r_{\rm max,TDE}$. We set $N_{\rm max}=4/3 \pi r_{\rm max,TDE}^3\rho_{\rm TDE} \approx 1$. 

With these parameters, we obtained $F_{\rm \nu,tot}^{\rm (TDE)} = 1.6\times10^{-8}$\,erg\,cm$^{-2}$\,s$^{-1}$. 

%%%%%%%%%%%%%%%%%%%%%%%%%%%%%%%%%%%%%%%%%%%%%%%%%%%%%%%%%
\subsubsection{Gamma-ray bursts}
%%%%%%%%%%%%%%%%%%%%%%%%%%%%%%%%%%%%%%%%%%%%%%%%%%%%%%%%%

While no significant neutrino emission has been associated with gamma-ray bursts (GRBs), here we considered the corresponding total flux from GRBs assuming $N_{\rm det}=2.3$, which is the 90\% confidence level upper limit corresponding to no detections. This can give us a picture of the total flux contribution that is consistent with the lack of association. We adopted a local density of $\rho_{\rm GRB} = \mathcal{R}_{\rm GRB}T_{\rm obs} = 10^{-8}$\,Mpc$^{-3}$ where we assumed $T_{\rm obs}=10$\,yr \citep{2010MNRAS.406.1944W}. We considered GRBs to be detectable with single-neutrino searches. 

Given the short duration and rarity of GRBs, and their easy observability with all-sky gamma-ray detectors, single-neutrino searches can consider neutrinos below the 100\,TeV limit used for other single-neutrino searches. Adopting a threshold of 1\,TeV and a neutrino spectral index of 3, we consider a threshold flux $\mathcal{F}_{\rm \nu,1}=4\times10^{-9}$\,erg\,cm$^{-2}$\,s$^{-1}(1\mbox{yr}/T_{\rm obs})$, i.e. 100 times lower than for an energy threshold of 100\,TeV. 

We adopted $f_0=2\times10^{-4}$\,Mpc$^{-1}$, similar to the fraction expected for a population tracing the star formation rate. We took $N_{\rm max}=1000$, comparable to the total number of detected GRBs. 

With these parameters, we obtained $F_{\rm \nu,tot}^{\rm (GRB)} \lesssim 10^{-9}$\,erg\,cm$^{-2}$\,s$^{-1}$. We conclude that the lack of detection presents a strong constraint on the GRB contribution to the total cosmic neutrino flux, their contribution is $<1\%$. This limit is consistent with constraints from searches by IceCube for neutrinos coincident with GRBs \citep{2017ApJ...843..112A}.

%%%%%%%%%%%%%%%%%%%%%%%%%%%%%%%%%%%%%%%%%%%%%%%%%%%%%%%%%
\subsubsection{Core-collapse supernovae}
%%%%%%%%%%%%%%%%%%%%%%%%%%%%%%%%%%%%%%%%%%%%%%%%%%%%%%%%%

No significant neutrino emission has been associated with supernovae. Here we considered the corresponding total flux from core-collapse supernovae assuming $N_{\rm det}=2.3$ to characterize the limit this lack of detection represents. 

We adopted a core-collapse supernova rate density of ${\sim}10^{-4}$Mpc$^{-3}$yr$^{-1}$ \citep{2014ApJ...792..135T}, and assumed that neutrino emission is significant for a duration of $T_{\rm obs}=10\,$yr. With this we obtain an effective density $\rho_{\rm SN}=10^{-3}$\,Mpc$^{-3}$. We considered supernovae to be detectable with time-integrated searches.  

We adopted $F_{\rm \nu,0}=2\times10^{-11}$\,erg\,cm$^{-2}$\,s$^{-1}$ where we assumed a neutrino spectral index of \add{2.5} and energy range $\varepsilon_\nu\in[100\mbox{GeV},100\mbox{PeV}]$ (see also \cite{2018PhRvD..97h1301M,2013PhRvL.111l1102M}). We further adopted \add{$f_0=1.5\times10^{-4}$\,Mpc$^{-1}$} assuming that the supernova rate traces the star formation rate.

With these parameters, we obtained $F_{\rm \nu,tot}^{\rm (CCSN)} \lesssim 10^{-7}$\,erg\,cm$^{-2}$\,s$^{-1}$, i.e. greater than IceCube's overall flux. We therefore conclude that the lack of observation does not present a meaningful constraint on the contribution of supernovae to the total cosmic neutrino flux. 

%%%%%%%%%%%%%%%%%%%%%%%%%%%%%%%%%%%%%%%%%%%%%%%%%%%%%%%%%
\subsubsection{\add{Starburst galaxies}}
%%%%%%%%%%%%%%%%%%%%%%%%%%%%%%%%%%%%%%%%%%%%%%%%%%%%%%%%%

\add{Starburst galaxies have been proposed as a major contributor to the cosmic high-energy neutrino flux \citep{2006JCAP...05..003L,2014JCAP...09..043T,2014PhRvD..89l7304A,2016PhRvD..94j3006M,2017ApJ...836...47B}. No direct association has been made so far. We adopted an effective number density of $\rho_{\rm starburst} = 10^{-5}$\,Mpc$^{-3}$ \citep{2016PhRvD..94j3006M}; $F_{\rm \nu,0}=2\times10^{-11}$\,erg\,cm$^{-2}$\,s$^{-1}$ where we assumed a neutrino spectral index of 2.5 and energy range $\varepsilon_\nu\in[100\mbox{GeV},100\mbox{PeV}]$; and $f_0=1.5\times10^{-4}$\,Mpc$^{-1}$ assuming that the starburst galaxies trace the star formation rate.}

\add{We further considered $N_{\rm det}=2.3$ to characterize the limit that the lack of detection represents. Using Eq. \ref{eq:ftotint} we obtained $F_{\rm \nu,tot}^{\rm (starburst)} \lesssim 2.5\times 10^{-8}$\,erg\,cm$^{-2}$\,s$^{-1}$, limiting the starburst contribution to $\lesssim 40\%$ of the IceCube flux. A higher effective number density for starburst galaxies would correspond to an even less stringent constraint.}

%%%%%%%%%%%%%%%%%%%%%%%%%%%%%%%%%%%%%%%%%%%%%%%%%%%%
\section{Full Bayesian model}
\label{sec:fullmodel}
%%%%%%%%%%%%%%%%%%%%%%%%%%%%%%%%%%%%%%%%%%%%%%%%%%%%

We now turn to a more detailed derivation of the expected neutrino flux from different source populations where we take into account the cosmic evolution of sources and the statistical uncertainty of the number of detections.

%%%%%%%%%%%%%%%%%%%%%%%%%%%%%%%%%%%%%%%%%%%%%%%%%%%%
\subsection{Probability density of the expected number of detections}
%%%%%%%%%%%%%%%%%%%%%%%%%%%%%%%%%%%%%%%%%%%%%%%%%%%%

Assume we have a set of $N_{\rm tot}$ detection candidates for source type $S$. Each candidate is either from the astrophysical source of interest or from the background. For our purposes here, astrophysical neutrinos from unassociated sources are counted as background. Candidate $i$ has a set of reconstructed parameters denoted with $\vec{x}_i$. The set of reconstructed parameters for all candidates is denoted with $\vec{x}=\{\vec{x}_1,\vec{x}_2...\vec{x}_{\rm N_{\rm tot}}\}$.

Let $p(\vec{x}_i|S)$ and $p(\vec{x}_i|B)$ be the probability densities of observing $\vec{x}_i$ from an astrophysical source of type $S$ and from the background, respectively. We compute the probability density of the expected number of detected events, denoted with $N_{\rm det}$ as \citep{2015PhRvD..91b3005F} (note the similarity to Eq. 7 in \citealt{2008APh....29..299B}):
\begin{equation}
\begin{split}
p(N_{\rm det} | \vec{x})\propto \int & dN_{\rm B} \prod_{i=1}^{N_{\rm tot}} \left[ N_{\rm det}p(\vec{x}_i|S) + N_{\rm B} p(\vec{x}_i|B)\right] \\ &\times e^{-(N_{\rm det} + N_{\rm B})}\frac{\pi(N_{\rm det})}{\sqrt{N_{\rm B}}}
\end{split}
\label{eq:p_theta}
\end{equation}
where $N_{\rm B}$ is the expected number of detected background events, which we marginalize over. We used the Poisson Jeffreys prior for $N_{\rm B}$. We define the prior probability density $\pi(N_{\rm det})$ of $N_{\rm det}$ implicitly by assuming that the prior probability density of the total flux from a given source type is uniformly distributed between 0 and IceCube's measured cosmic neutrino flux minus the flux of the other sources. Therefore, we will have a three-dimensional prior probability density for the three sources considered below, with uniform probability density and with the boundary condition that the sum of the total flux from the three sources cannot exceed IceCube's measured flux.

%%%%%%%%%%%%%%%%%%%%%%%%%%%%%%%%%%%%%%%%%%%%%%%%%%%%
\subsection{Computing the neutrino luminosity of individual sources}
%%%%%%%%%%%%%%%%%%%%%%%%%%%%%%%%%%%%%%%%%%%%%%%%%%%%

If we know the expected number of detected sources, we can compute the related neutrino luminosity of individual sources. This computation, however, also depends on the luminosity's probability density. 

Here we will assume that the neutrino luminosity $L_\nu$ of a source depends on its electromagnetic luminosity $L_\gamma$ \addd{(see Section \ref{sec:implementation} for model-dependent assumptions on $L_\gamma$)}. For simplicity, we will assume that $L_\nu = \alpha_{\gamma\nu} L_\gamma$ with unknown $\alpha_{\gamma\nu}$ constant. We further assume that we know the number density $\rho(z,L_\gamma)$ of a continuous source type as a function of redshift $z$ and $L_\gamma$. 

With this, we can compute the expected number of detections (see also \citealt{2016PhRvD..94j3006M}):
\begin{equation}
N_{\rm S,cont.} = \int  dz dL_{\nu} \frac{4\pi c}{H(z)} \frac{d_{\rm L}(z)^2}{(1+z)^2} \rho(z,\frac{L_{\nu}}{\alpha_{\gamma\nu}}) p_{\rm det}(z,L_{\nu})\,.
\label{eq:NScont}
\end{equation}
Here, $d_{\rm L}$ is the luminosity distance and $p_{\rm det}(z,L_{\nu})$ is the probability that a source with luminosity $L_{\nu}$ at redshift $z$ will be detected. We can compute the expected detection rate for transient sources as well, where we also need to take into account  time dilation, obtaining 
\begin{equation}
\begin{split}
N_{\rm S,trans.} = T_{\rm obs} \int  dz dE_{\nu} \frac{4\pi c}{H(z)} \frac{d_{\rm L}(z)^2}{(1+z)^3}  \\ \times \mathcal{R}(z,\frac{E_{\nu}}{\alpha_{\gamma\nu}}) p_{\rm det}(z,E_{\nu})\,.
\end{split}
\label{eq:NStrans}
\end{equation}
Here, $T_{\rm obs}$ is the duration of observation, $\mathcal{R}(z,E_{\gamma})$ is the comoving rate density and $E_{\gamma}=\alpha_{\gamma\nu}E_\nu$ is the radiated electromagnetic energy.

We can then compute the unknown $\alpha_{\gamma\nu}$ factor by equating the expected number of detections from Eqs. \ref{eq:NScont} or \ref{eq:NStrans} with the expected number of detections from observations.

Note that if we assumed that all sources have the same neutrino luminosity then this step would not be needed, and we could simply compute the overall neutrino flux from the number of detections from observations. However, this step enables the incorporation of the source luminosity distribution, which we can base on the observed $\gamma$ luminosity distributions for the source types in question.

\add{An interesting side product of this step is the determination of $\alpha_{\gamma\nu}$, i.e. the connection between the sources' gamma flux and expected neutrino flux given the number of detections. For blazars and AGNs discussed below, we obtain a characteristic conversion  factor of $\alpha_{\gamma\nu}^{\rm blazar}\sim 5$ and $\alpha_{\gamma\nu}^{\rm AGN}\sim 0.04$. These characteristic values are obtained by assuming that the number of detections is the expected number (as considered in our "simple model"). It therefore appears that blazars are more efficient neutrino producers than AGNs.}

%%%%%%%%%%%%%%%%%%%%%%%%%%%%%%%%%%%%%%%%%%%%%%%%%%%%
\subsection{Expected total flux at Earth}
%%%%%%%%%%%%%%%%%%%%%%%%%%%%%%%%%%%%%%%%%%%%%%%%%%%%

To obtain the expected total neutrino flux at Earth for a source type, we integrate over all sources in the universe. We also marginalize over the distribution of the expected number of detections from Eq. \ref{eq:p_theta}. For continuous sources we obtain

\begin{equation}
\begin{split}
\mathcal{F_{\nu,{\rm S,cont.}}} = \int  dN_{\rm det}dz dL_{\nu}d\theta \frac{c}{H(z)} \rho(z,L_\nu,N_{\rm det}) \\ \times p_{\rm det}(z,L_{\nu})\frac{L_{\nu}}{(1+z)^{2+\alpha}}p(N_{\rm det} | \vec{x})\,.
\end{split}
\label{eq:totalfluxcont}
\end{equation}
where $\alpha$ is the spectral index of the neutrino spectral density $dN_\nu/d\varepsilon_\nu\propto \varepsilon^{-\alpha}$, and we expressed $\rho$ as a function of $N_{\rm det}$ and $L_{\nu}$. This latter takes into account that we do not know the density spectrum of $L_{\nu}$, but can determine it using the density spectrum of $L_\gamma$ and $N_{\rm det}$, and by assuming that $L_{\nu}\propto L_\gamma$. We can similarly compute the total flux at Earth for transients:
\begin{equation}
\begin{split}
\mathcal{F_{\nu,{\rm S,trans.}}} = \int  dN_{\rm det}dz dL_{\nu}d\theta \frac{c}{H(z)} \mathcal{R}(z,L_\nu,N_{\rm det}) \\ \times p_{\rm det}(z,L_{\nu})\frac{L_{\nu}}{(1+z)^{3+\alpha}}p(N_{\rm det} | \vec{x})\,.
\end{split}
\label{eq:totalfluxtran}
\end{equation}

%%%%%%%%%%%%%%%%%%%%%%%%%%%%%%%%%%%%%%%%%%%%%%%%%%%%
\section{Implementation}
\label{sec:implementation}
%%%%%%%%%%%%%%%%%%%%%%%%%%%%%%%%%%%%%%%%%%%%%%%%%%%%

Here we implement the general framework discussed above by considering available information from observations.

%%%%%%%%%%%%%%%%%%%%%%%%%%%%%%%%%%%%%%%%%%%%%%%%%%%%
\subsection{Source parameter probability densities}
%%%%%%%%%%%%%%%%%%%%%%%%%%%%%%%%%%%%%%%%%%%%%%%%%%%%

We now turn to $p(\vec{x}_i|B)$ and $p(\vec{x}_i|S)$. 
We consider $\vec{x}_i=p_i$. The probability density of the $p$-value is naturally defined for the background distribution as uniform ($p(\vec{x}_i|B) = 1$). However, the signal distribution $p(\vec{x}_i|z_i,L_{\nu,i}S)$ cannot be determined without a specific astrophysical signal model and the data analysis framework used in the search. Instead, we adopt a "calibrated" ratio of the background and signal probability densities from \cite{pvalue}:
\begin{equation}
\frac{p(\vec{x}_i|B)}{p(\vec{x}_i|S)} = -e \,p_i \ln(p_i)
\label{eq:ratio}
\end{equation}
for $p_i<1/e$, where $e$ is Euler's number. This ratio is a lower bound over a wide range of realistic $p$-value distributions for the signal hypothesis \add{where the only assumption is that the density of $p_i$ under the signal hypothesis should be decreasing in $p_i$}. \addd{As this "calibrated" ratio is a lower bound it may still be optimistic in terms of the flux contributions, nevertheless we consider this a reasonable "calibrated" estimate given the unknown signal hypothesis.}

%%%%%%%%%%%%%%%%%%%%%%%%%%%%%%%%%%%%%%%%%%%%%%%%%%%%
\subsection{Detection probability}
%%%%%%%%%%%%%%%%%%%%%%%%%%%%%%%%%%%%%%%%%%%%%%%%%%%%

For a given redshift and neutrino luminosity, the probability of detection depends on both the identification of the source through electromagnetic observations and on the detectability of the neutrino signal. Electromagnetic identification determines the completeness $\mathcal{C}(z, L_\gamma)$ of a source catalog, where $L_\gamma$ is the electromagnetic luminosity of the source. We assumed here that neutrino luminosity is proportional to the electromagnetic luminosity of the source, i.e. $L_\gamma\propto L_\nu$ 

The detectability of the neutrino signal depends on multiple factors, including the neutrino flux and spectrum at Earth, detector sensitivity, and the number density of the source population. In the limit of rare sources, even a single neutrino will be sufficient to identify a source. In this case, detectability will be the probability that a single neutrino is recorded. To obtain this probability for a given source flux, we adopted the expected number of detected neutrinos for a given flux presented by \cite{AMONEHEalert}. For a source spectrum $dN_{\nu}/d\varepsilon_{\nu}\propto\varepsilon_{\nu}^{-2.5}$, the flux density corresponding to an expected 1 detected neutrino is $1.2\times10^{-17}$\,GeV$^{-1}$\,cm$^{-2}$\,s$^{-1}(\varepsilon_\nu / 100\,\mbox{TeV})^{-2.5}(T_{\rm obs}/\mbox{1\,yr})^{-1}$. 

For more common sources, detectability can be approximated with a flux threshold $\mathcal{F}_{\nu,th}$ such that all those -- and only those -- sources with flux $\mathcal{F} \geq \mathcal{F}_{\nu,th}$ are detected. Below we adopt $\mathcal{F}_{\nu,th}$ from IceCube's 10-year, 90\% confidence-level median sensitivity based on \cite{2020PhRvL.124e1103A}, integrated within $[100\,\mbox{GeV},100\,\mbox{PeV}]$. For $E^{-3}$ and $E^{-2}$ spectra this sensitivity is $\mathcal{F}_{\nu,th,3}=2\times10^{-10}$\,erg\,cm$^{-2}$\,s$^{-1}$ and $\mathcal{F}_{\nu,th,2}=10^{-11}$\,erg\,cm$^{-2}$\,s$^{-1}$, respectively. To obtain the sensitivity for an $E^{-2.5}$ spectrum, we consider the geometric mean of $\mathcal{F}_{\nu,th,3}$ and $\mathcal{F}_{\nu,th,2}$, obtaining $\mathcal{F}_{\nu,th,2.5}=2\times10^{-11}$\,erg\,cm$^{-2}$\,s$^{-1}$. These values are valid for the northern hemisphere. We account for the fact that IceCube is much more sensitive towards the northern hemisphere by introducing an effective factor of 2 reduction in source completeness.

\add{To characterize the dependence of our results to these approximate sensitivity threshold, we can look at Eqs. \ref{eq:ftotint} \& \ref{eq:ftotsingle}, which show in the case of our simplified model that the results are linearly dependent on the thresholds. This can be somewhat mitigated in our full model by the overall constraint that the total flux is less than the IceCube flux. }

%%%%%%%%%%%%%%%%%%%%%%%%%%%%%%%%%%%%%%%%%%%%%%%%%%%%
\subsection{Source types}
%%%%%%%%%%%%%%%%%%%%%%%%%%%%%%%%%%%%%%%%%%%%%%%%%%%%
Here we introduce the source properties used for the analysis.

\begin{table}
\begin{center}
\hspace{-5mm}
\begin{tabular}{llll}
\hline 
Name & Type & $p$ & Ref. \\
\hline\hline
NGC\,1068 & AGN & $0.008$ & \cite{2020PhRvL.124e1103A} \\
TXS\,0506+056 & blazar & $0.001$ & \cite{2018Sci...361.1378I}\\
PKS\,1502+106 & blazar & $0.01$ & \cite{2019ATel12967....1T} \\
PKS\,1424-41 & blazar & $0.05$ & \cite{2016NatPh..12..807K}\\
AT2019dsg & TDE & $0.002$ & \cite{2021NatAs...5..510S}\\
\hline
\end{tabular}
\caption{\add{Astrophysical sources with associated neutrino emission used in this analysis.} $p$ is the $p$-value of the neutrino signal's association with the astrophysical source.}
\label{table:sources}
\end{center}
\end{table}

\subsubsection{Active galactic nuclei}

One active galactic nucleus (AGN), NGC 1068, has been identified as a likely neutrino source at 2.9$\sigma$ post-trial significance \cite{2020PhRvL.124e1103A}. We use this one detection, with p-value $p=2\times 10^{-3}$. We adopt the cosmic number density and luminosity function $\rho_{\rm AGN}(z,L_\gamma)$ for AGNs from the Spitzer mid-infrared AGN survey \cite{2015ApJ...802..102L}, with threshold $L_\gamma\geq10^{41}$\,erg\,s$^{-1}$. We consider AGNs detected as neutrino sources if their neutrino flux is above the threshold $\mathcal{F}_{\nu,th}$. This flux threshold is adopted based on the measured (90\% C.L. median) sensitivity of IceCube's 10-year search \citep{2020PhRvL.124e1103A}. We use the typical sensitivity at the northern hemisphere given that IceCube is much more sensitive in this direction, and adopt a factor of 2 reduction in the completeness of the AGN catalog. For a source with neutrino spectral index 2, this corresponds to $\mathcal{F}_{\nu,th,2}=6\times10^{-13}$\,erg\,cm$^{-2}$\,s$^{-1}$ considering neutrino energies $\varepsilon_\nu\in[100\mbox{GeV},100\mbox{PeV}]$. For neutrino spectral index 3 we find $\mathcal{F}_{\nu,th,3}=2\times10^{-10}$\,erg\,cm$^{-2}$\,s$^{-1}$.

In the detection process, only a few AGNs have been used in neutrino searches, selected based on their $\gamma$-ray brightness \citep{2020PhRvL.124e1103A}. This limitation ensured that the trial factor in the search remained low. We take this into account by limiting the completeness of our simulated catalog to the 10 brightest sources on the northern hemisphere, i.e. $p_{\rm det}$ is set to zero for sources with $z$ and $L_\gamma$ for which the electromagnetic flux at earth is below a threshold ($10^{-9}$\,erg\,cm$^{-2}$\,s$^{-1}$) such that 10 sources are expected. 

\subsubsection{Blazars}

For blazars we consider 3 detections (see Table \ref{table:sources}). We adopt the cosmic number density and radio luminosity function $\rho_{\rm blazar}(z,L_\gamma)$ for FSRQs from \cite{2017ApJ...842...87M}, with threshold $L_\gamma\geq10^{40}$\,erg\,s$^{-1}$. We consider a blazar detected if a single extremely high energy neutrino with energy above 100\,TeV is detected from it. For a source with spectral index 2, one extremely high energy neutrino is expected to be detected from a source in a random direction \citep{AMONEHEalert} for a neutrino flux of $\mathcal{F}_{\rm \nu,single,2}=4\times10^{-10}$\,erg\,cm$^{-2}$\,s$^{-1}(1\mbox{yr}/T_{\rm obs})$ in the neutrino energy range range $\varepsilon_\nu\in[100\mbox{TeV},100\mbox{PeV}]$. For spectral index 3 we also have $\mathcal{F}_{\rm \nu,single,3}=4\times10^{-10}$\,erg\,cm$^{-2}$\,s$^{-1}(1\mbox{yr}/T_{\rm obs})$.

Similarly to AGNs, we limit the number of blazars in our search catalog. IceCube's 10-year catalog search included about 100 blazars \citep{2020PhRvL.124e1103A}, we therefore adopt the same number here.

\subsubsection{TDEs}

The detection of one TDE, AT2019dsg, has been reported so far in \cite{2021NatAs...5..510S}, which we include in this analysis. We adopt the cosmic rate density of TDEs from \cite{2015ApJ...812...33S}, using a minimum TDE luminosity of \addd{$L_\gamma=10^{44}$\,erg\,s$^{-1}$}. \add{We consider the rate and evolution of all TDEs, i.e. we do not require the presence of TDE jets.} For simplicity, we treat all TDEs as having identical neutrino luminosity and detectability. 

Given the regular electromagnetic follow-up of very-high-energy neutrinos released publicly by IceCube, we will consider the catalog of TDEs complete out to the distance they can be found through electromagnetic observations. Given the distance range of past identified TDEs \citep{2020arXiv200101409V}, we set their detectable distance at $r_{\rm max,TDE}=1$\,Gpc.

We can compute the total fluence needed for the expected detection of a single neutrino analogously for blazars, but using fluence instead of flux. For a source with spectral index 2, we find a neutrino fluence of $\mathcal{S}_{\rm \nu,single,2}=0.1$\,erg\,cm$^{-2}$ in the neutrino energy range  $\varepsilon_\nu\in[100\mbox{TeV},100\mbox{PeV}]$. For spectral index 3 we also have $\mathcal{S}_{\rm \nu,single,3}=0.1$\,erg\,cm$^{-2}$.

%%%%%%%%%%%%%%%%%%%%%%%%%%%%%%%%%%%%%%%%%%%%%%%%%%%%
%\section{Results}
%%%%%%%%%%%%%%%%%%%%%%%%%%%%%%%%%%%%%%%%%%%%%%%%%%%%

\begin{figure}
\includegraphics[width=0.48\textwidth]{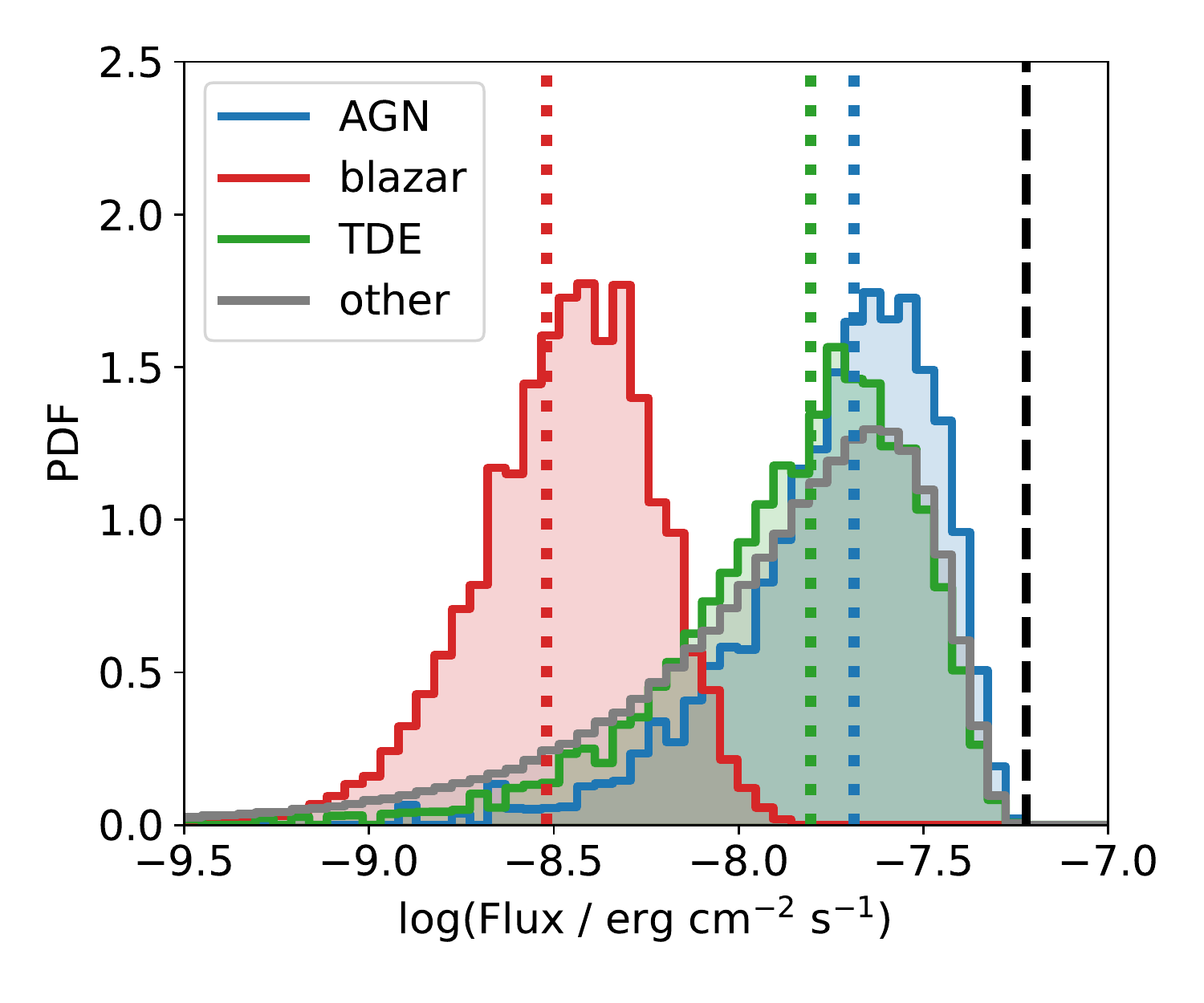}
\centering
\caption{{\bf Probability density of total neutrino flux} from AGNs, blazars and TDEs based on IceCube's detected sources. For comparison we show the total measured IceCube flux (vertical dashed black line; \citealt{2020arXiv200109520I}), and the estimated flux for the three source types using our simple model (Section \ref{sec:simplemodel}; vertical dotted lines). We assumed an $E^{-2.5}$ astrophysical neutrino spectrum \citep{2020arXiv200109520I}. 
}
\label{fig:fluxdist}
\end{figure}

\begin{figure}
\includegraphics[width=0.5\textwidth]{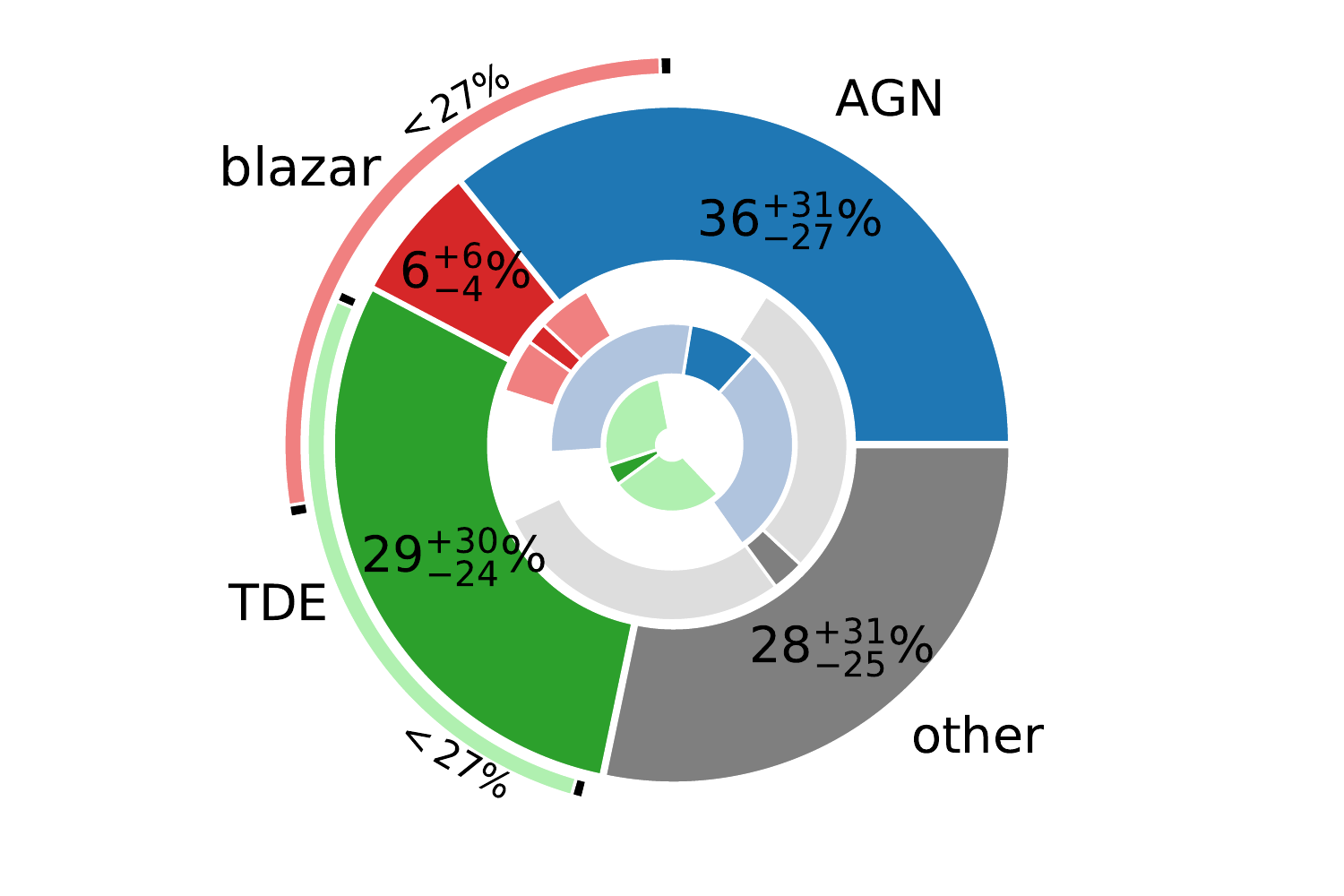}
\centering
\caption{\add{{\bf Pie chart of expected fraction of IceCube's neutrino flux} from the three source types with directly associated neutrino detections, and the remaining flux from unknown sources (other). The main pie chart indicates the average expected fractional values, while the innermost charts indicate  uncertainty by showing the minimum (dark) and maximum (light) extent of the fractional contributions allowed within the 90\% \addd{credible} regions. Outer thin slices further show independent observational constraints from stacking analyses for TDEs \citep{2019ICRC...36.1016S} and blazars \citep{Aartsen_2017} that limit their contributions to $<27\%$ each at 90\% confidence level.}
}
\label{fig:piechart}
\end{figure}

\begin{center}
\begin{table}
\hspace{-5mm}
\begin{tabular}{|c|c|c|c|}
\hline 
%Source type & Flux [$10^{-7}$\,erg\,s$^{-1}$\,cm$^{-2}$] \\
\multirow{2}{*}{\bf{Type}} &\multicolumn{3}{c|}{Flux  / $\phi_{\rm IC}$} \\ \cline{2-4}
 
& warm-up & simple & full \\
\hline\hline
%AGN    & & & $2.5^{+2.1}_{-1.9}$ \\
%blazar & & & $0.9^{+0.8}_{-0.6}$ \\
%TDE    & & & $2.4^{+2.3}_{-1.9}$ \\
%other  & & & $1.6^{+2.1}_{-1.4}$ \\
AGN       &      & 0.34    & $0.36^{+0.31}_{-0.27}$ \\
blazar    & 0.1  & 0.05     & $0.06^{+0.06}_{-0.04}$ \\
TDE       & 0.45 & 0.26    & $0.29^{+0.30}_{-0.24}$ \\
other     &      &         & $0.28^{+0.38}_{-0.25}$ \\
\hline
GRB       &      & $<0.01$ &  \\
CCSN      &      & $<1$  &  \\
\add{starburst} &      & \add{$<0.4$}  &  \\
\hline
\end{tabular}
\caption{Estimated cosmic neutrino flux as a fraction of IceCube's total measured flux ($\phi_{\rm IC}$). Results are shown for the "warm-up" (Section \ref{sec:warmup}), "simple" (Section \ref{sec:simplemodel}) and "full" Bayesian (Section \ref{sec:fullmodel}) models, for AGNs, blazars and TDEs, and the total estimated flux from unknown source types. Error bars indicate 90\% credible interval. For our "simple" model we also show upper limits for GRBs, indicating that non-detection presents are very strict constraint on their allowed contribution to the overall flux due to their very low rate density, and core-collapse supernovae (CCSNe), showing that non-detection does not meaningfully constrain their contribution to the overall neutrino flux due to their high rate density.}
\label{table:flux}
\end{table}
\end{center}

%%%%%%%%%%%%%%%%%%%%%%%%%%%%%%%%%%%%%%%%%%%%%%%%%%%%
\section{Results}
\label{sec:results}
%%%%%%%%%%%%%%%%%%%%%%%%%%%%%%%%%%%%%%%%%%%%%%%%%%%%

We computed the expected cosmic flux for AGNs, blazars, and TDEs using the above prescription. To understand the statistical uncertainty of our results, we obtained the flux probability densities for the three cases based on the probability density $p(N_{\rm det} | \vec{x})$ in Eq. \ref{eq:p_theta}. Flux probability densities are computed similarly to Eq. \ref{eq:totalfluxcont} for AGNs and blazars, and Eq. \ref{eq:totalfluxtran} for TDEs, but without marginalization over $N_{\rm det}$. \addd{Specifically, in this step we compute the probability densities $\partial\mathcal{F_{\nu,{\rm S,cont.}}}/\partial N_{\rm det}$ and $\partial\mathcal{F_{\nu,{\rm S,trans.}}}/\partial N_{\rm det}$ by carrying out the source simulation with different $N_{\rm det}$ values and then converting the array of results into a distribution.} 

The results are shown in Fig. \ref{fig:fluxdist}. We also list the expected values and 90\% credible intervals in Table \ref{table:flux}. We see that AGNs and TDEs have comparable expected fluxes, while the expected flux from blazars is about a factor of 3 lower than these. We also see that, due to the low number of detections so far, the expected flux has considerable uncertainties. 

To obtain a pie chart of total fluxes, we look at the properties of the cosmic quasi-diffuse neutrino flux. IceCube detections are consistent with an astrophysical flux following a power-law distribution with spectral index of $2.53\pm0.07$ \cite{2020arXiv200109520I}. Therefore, we consider a neutrino spectrum that scales as $E^{-2.5}$ with neutrino energy $E$. Our results would be similar to those presented below if we adopted a somewhat softer spectrum with $E^{-3}$ that best fits the energy distribution of the highest energy neutrinos \citep{2020arXiv201103545A}.

We combine our results together for blazars, AGNs, and TDEs into a {\it pie chart} that shows the expected relative abundance of these three source types in the overall flux of high-energy neutrinos IceCube is detecting. The obtained pie chart is shown in Fig. \ref{fig:piechart}. 

\subsection{Expected contribution from other source types}

Based on the probability densities of AGNs, blazars and TDEs shown in Fig. \ref{fig:fluxdist}, we computed the expected contribution from other (unspecified) source types. For this we considered the AGN, blazar, and TDE probability densities to be independent and computed the probability density of their combined neutrino flux, with the boundary condition that the total cannot exceed IceCube's measured total flux. We found that the unknown sources represent at least 10\% (1\%) of IceCube's total flux with 80\% (98\%) probability. \addd{This fraction could be even higher given our (upper bound) estimation of the probability density ratio in Eq. \ref{eq:ratio}.}

\add{We note that this computation only accounts for the observed associations. It does not take into account independent source constraints or theoretically expected source energetics \citep{2019PhRvD..99f3012M} or emission efficiencies \citep{2020PhRvL.125a1101M}. It is therefore interesting to compare our results with alternative expectations, which are largely consistent with our results within uncertainties.}

%%%%%%%%%%%%%%%%%%%%%%%%%%%%%%%%%%%%%%%%%%%%%%%%%%%%
\section{\add{Independent observational limits}}
\label{sec:limits}
%%%%%%%%%%%%%%%%%%%%%%%%%%%%%%%%%%%%%%%%%%%%%%%%%%%%

\add{While the present work computes fractional source contributions from associated events, source contribution limits have been previously derived through independent observing strategies.}

\add{A main strategy is to use the non-detection of very-high energy neutrino multiplets to constrain the the flux of different source types \citep{2015JPhCS.632a2039K,2016PhRvD..94j3006M,2020PhRvD.101l3017C}. The lack of such multiplets particularly limits transients, such as GRBs and TDEs, and rare source types, such as blazars. These source types are excluded as the dominant sources of the observed quasi-diffuse neutrino flux.}

\add{A stacking analysis found that TDEs cannot contribute more than 27\% of the total diffuse astrophysical neutrino flux at 90\% confidence level \citep{2019ICRC...36.1016S}.}

\add{The contribution of known blazars (those in Fermi's second catalog) to the total neutrino flux within 10\,TeV and 2\,PeV was limited by a stacking search to less than 27\% assuming a neutrino spectrum with index $-2.5$ \citep{Aartsen_2017}.}

\add{Stacked searches have also been carried out for Type Ibc core-collapse supernovae \citep{2018JCAP...01..025S,2018JCAP...12..008E}, but these do not rule out these sources as major contributors to the IceCube flux.}

%%%%%%%%%%%%%%%%%%%%%%%%%%%%%%%%%%%%%%%%%%%%%%%%%%%%
\section{Conclusion}
\label{sec:conclusion}
%%%%%%%%%%%%%%%%%%%%%%%%%%%%%%%%%%%%%%%%%%%%%%%%%%%%

We computed the expected total high-energy neutrino flux at Earth from AGNs, blazars, and TDEs based on the associations of individual sources with astrophysical neutrinos detected by IceCube, IceCube's sensitivity, and the astrophysical properties and distributions of the three source types. We first carried out a simple derivation of the expected neutrino flux in order to demonstrate how the results scale with the properties of the detections, IceCube and the sources. We then carried out a more detailed derivation that accounts for the statistical uncertainty of the detection process, varying neutrino luminosity within a source type, and the cosmic evolution of source densities and properties. Our conclusions are as follows:
\begin{itemize}[leftmargin=*,itemsep=1pt,topsep=1pt]
\item Despite having detected more blazars with neutrinos than AGNs or TDEs, blazars are expected to be the smallest contributor to the cosmic neutrino flux. We found their contribution to be $3.9^{+3.7}_{-2.6} \times 10^{-9}$\,erg\,s$^{-1}$\,cm$^{-2}$ (error bars indicate 90\% credible interval), or a contribution that is $<11\%$ of the total quasi-diffuse neutrino flux detected by IceCube (at 90\% credible level). This relatively small contribution is due to the fact that blazars are rare, making them much easier to identify through multi-messenger searches than a more common source type with similar total flux contribution. Significant contribution from low-luminosity blazars, nevertheless, could increase their fractional contribution \citep{Palladino_2019}.
\item AGNs and TDEs represent similar overall contributions. We estimated the AGN flux to be $2.1^{+1.8}_{-1.6} \times 10^{-8}$\,erg\,s$^{-1}$\,cm$^{-2}$, while for TDEs the estimated flux is $1.8^{+1.8}_{-1.4} \times 10^{-8}$\,erg\,s$^{-1}$\,cm$^{-2}$. Either AGNs or TDEs could be the majority source of cosmic high-energy neutrinos. Their most likely contribution is about $1/3^{\rm rd}$ of the total flux for each.
\item One or more, so far unidentified, source types also likely contribute to the overall flux. We estimate their contribution to be $1.7^{+2.3}_{-1.5} \times 10^{-8}$\,erg\,s$^{-1}$\,cm$^{-2}$. 
\end{itemize}

\add{The above results only accounted for information in source types with neutrino associations, i.e. our pie chart does not consider the breakdown of "other" source types, or a more detailed prior theoretical expectations from promising source types such as starburst galaxies, galaxy clusters or supernovae. We also do not fold in information from observational limits from stacked or other searches. Further sources of uncertainty include the assumed neutrino spectrum, which may have a different, and possibly not-power-law, spectrum for different source types \citep{2018npa..confE..84P}. Despite these limitations, our results are broadly consistent with other observational and theoretical expectations within uncertainties.}

The diversity of neutrino sources that apparently contribute to the diffuse flux and the further possibility of unidentified classes of neutrino sources that remained hidden so far makes future observations, and next generation neutrino observatories such as IceCube-Gen2 \citep{2020arXiv200804323T} or KM3NeT \citep{2016JPhG...43h4001A}, particularly interesting. The discovery of more AGNs, blazars and TDEs, and the better resolution of the astrophysical high-energy neutrino spectrum, will also be key in enabling the characterization of unidentified source types even if they remain undetected through electromagnetic observations.

% -------------------------------------------
% \section*{Acknowledgments}
% -------------------------------------------
\begin{acknowledgments}

The authors thank Cecilia Lunardini, Kohta Murase, Foteini Oikonomou, Christopher Wiebusch and Walter Winter for their valuable suggestions. The authors are grateful for the generous support of the University of Florida, DESY, and Columbia University in the City of New York.
I.B. acknowledges the support of the National Science Foundation under grant PHY-1911796 and the Alfred P. Sloan Foundation.

\end{acknowledgments}

\bibliography{Refs}

\end{document}